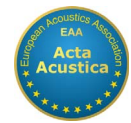

Available online at:
https://acta-acustica.edpsciences.org

SCIENTIFIC ARTICLE

**OPEN ∂ ACCESS**

# Auralization based on multi-perspective ambisonic room impulse responses

Kaspar Müller and Franz Zotter*

Institute of Electronic Music and Acoustics, University of Music and Performing Arts Graz, 8010 Graz, Austria

**Abstract** – Most often, virtual acoustic rendering employs real-time updated room acoustic simulations to accomplish auralization for a variable listener perspective. As an alternative, we propose and test a technique to interpolate room impulse responses, specifically Ambisonic room impulse responses (ARIRs) available at a grid of spatially distributed receiver perspectives, measured or simulated in a desired acoustic environment. In particular, we extrapolate a triplet of neighboring ARIRs to the variable listener perspective, preceding their linear interpolation. The extrapolation is achieved by decomposing each ARIR into localized sound events and re-assigning their direction, time, and level to what could be observed at the listener perspective, with as much temporal, directional, and perspective context as possible. We propose to undertake this decomposition in two levels: Peaks in the early ARIRs are decomposed into jointly localized sound events, based on time differences of arrival observed in either an ARIR triplet, or all ARIRs observing the direct sound. Sound events that could not be jointly localized are treated as residuals whose less precise localization utilizes direction-of-arrival detection and the estimated time of arrival. For the interpolated rendering, suitable parameter settings are found by evaluating the proposed method in a listening experiment, using both measured and simulated ARIR data sets, under static and time-varying conditions.

## 1 Introduction

An interactive, variable listener perspective in virtual acoustic environments necessitates rendering of movements in six degrees of freedom (6DoF), i.e. auralization for arbitrary translation and orientation of a listener. Utilizing room simulations is a common means to achieve variable-perspective audio rendering [1–3]. Naturally, such auralizations require a certain level of sophistication to reach authenticity even in static scenarios, and they should moreover ensure smooth transitions when the listener position is time-variant. A comparison of recent room simulation algorithms revealed that auralizations are perceived to be mostly plausible, although not authentic [3]. This motivates the development of approaches based on multi-perspective recordings or measured natural acoustic environments as alternatives, and the development of perspective interpolation approaches as helpful simplifications.

There are recent works that extrapolate single-perspective, first or higher-order Ambisonic recordings by projecting directionally localized sound objects onto a outer convex hull or onto predefined virtual room walls to achieve auralization in 6DoF [4–8]. Other approaches use plane-wave translations for variable-perspective auralization of single Ambisonic recordings [9, 10]. However, these methods mostly become inaccurate for great extrapolation distances.

To enable position shifts in a wider area, one can consider interpolating between Ambisonic recordings captured at multiple perspectives in the room simultaneously. While basic approaches apply a distance-weighted linear interpolation of the recorded perspectives near the listener [11, 12], parametric methods mostly use spatial time-frequency processing to extract and localize sound sources in order to synthesize them at a desired listener perspective [13–16], mixed with diffuse or unlocalized sound field residuals. A detailed overview of existing methods is provided in [17]. To avoid typical artifacts of signal-dependent time-frequency filtering, some other works render audio in 6DoF using broadband processing [18–21], which, however, would stay limited in spatial precision.

Auralization by convolution with measured multi-perspective directional room impulse responses (DRIRs) is an alternative to the rendering of multi-perspective surround recordings. Such a convolution approach complies with typical auralization, in which the directional impulse response is obtained from room-acoustics simulation. Despite a measured multi-perspective DRIR grid is expected to be much coarser, it is interchangeable with simulation to some degree. In contrast to Ambisonic or surround recordings, a convolution approach allows to auralize any single-channel signal in the desired spatial environment. And most importantly, measured DRIRs provide better accessibility to parametric decomposition of the captured acoustic environment, when compared to recorded signals. In particular, parametric decomposition is simplified by the temporal and directional sparsity of the sound events arriving in the early DRIR. Consequently, multi-perspective DRIRs enable a more detailed morphing of sound events observed at the recording perspective towards such observed at another, extrapolated perspective. The availability of multiple perspectives allows to superimpose their perspective extrapolation for interpolation.

---

*Corresponding author: zotter@iem.at

As an example, binaural room impulse responses (BRIRs) are commonly used as DRIRs to interface with headphone playback. Some variable-perspective rendering methods work with cross-fading between spatially distributed BRIRs [22, 23]. More sophisticated decomposition algorithms employ dynamic time warping [24–26]. However, fixed-orientation BRIRs do not provide an easy implementation of dynamic head rotation, and thus 6DoF approaches usually require to be fed with dummy head measurements done in multiple head orientations for every perspective [27, 28], which is exhaustive in terms of data and measurement effort. An approach described in [29] reduces the effort to a single-perspective BRIR measurement, however with multiple source directions and distances.

Alternatively to auralization from interpolated BRIRs, auralization can also be based on convolution with Ambisonic room impulse responses (ARIRs) as DRIRs. Other than with BRIRs, the convolution with ARIRs generates Ambisonic signals that are freely rotatable and can be decoded to both loudspeaker arrays and headphones. For single perspectives, there are multiple works presenting an efficient parameterization and directional enhancement of first-order Ambisonic room impulse responses using SIRR, SDM, ASDM [30–33]. Hence, auralization of translatory movements based on spatially distributed ARIRs is a promising approach for variable-perspective audio rendering. A suitable set of multi-perspective DRIRs could consist of *B*-format RIRs [34], i.e. first-order ARIRs.

Clearly, the known variable-perspective rendering methods [11–16, 18] could be applied to multi-perspective ARIRs. For instance, a distance-weighted, linear interpolation of the ARIRs closest to the desired listener perspective [35] is probably the most basic approach for variable-perspective rendering. However, as this and most of the known more sophisticated methods are designed to interpolate recorded Ambisonic signals rather than ARIRs, we expect an unused potential for a strong increase in spatial definition. This is because ARIR interpolation permits to employ methods that take advantage of the temporal sparsity and spatial localizability in the early ARIR parts. In particular, spatial resolution could be vastly improved when extracting and localizing high-energy sound events by incorporating temporal and directional information that is contained in multi-perspective ARIRs. The works [24, 25] describe an interpolation approach for first-order ARIRs based on dynamic time warping (DTW) that ensures temporal matching of the interpolated ARIRs. However, DTW does not enforce a geometrically consistent mapping of times, levels, and directions yet. A more elaborated method derived from DTW employs first-order ARIR interpolation including peak detection and matching, and a separate interpolation of directions of arrival [36].

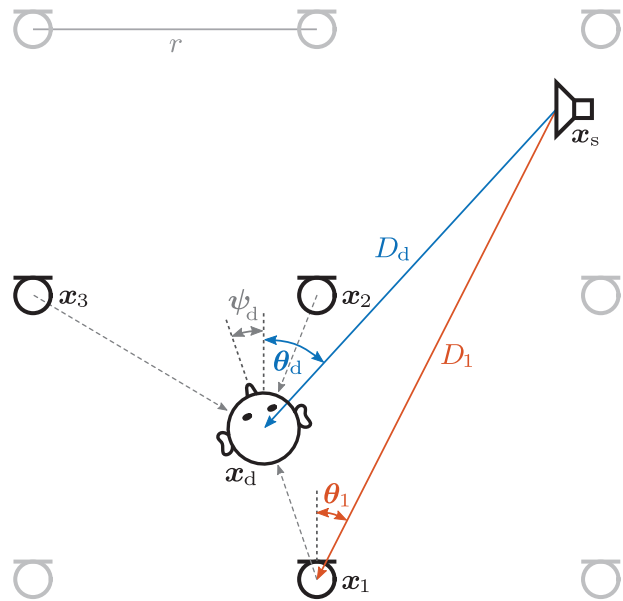

**Figure 1.** Rendering of a variable listener perspective by interpolation of the three closest ARIR perspectives of a spatially distributed ARIR grid. Before interpolation, the three ARIRs are extrapolated to the desired listener perspective $x_\mathrm{d}$ by decomposition into localized sound events, which are reproduced at the listener perspective with reassigned direction, time, and level.

To moreover ensure geometrical consistency exploiting the information available, our contribution introduces perspective interpolation from a spatially distributed ARIR triplet in Section 2. It employs extrapolation of the three ARIR perspectives based on localized instantaneous sound events, and it linearly interpolates the three ARIRs subsequently (cf. Fig. 1). Furthermore, we review the ASDM technique for directional resolution enhancement of first-order ARIRs by upmixing to higher Ambisonic orders, as reasonable extension of the ARIR interpolation. In Section 3, we propose an approach to extrapolate single ARIRs that restores temporal context within the ARIR using a simplistic sound-event localization and resampling technique. Section 4 introduces a more contextual, joint localization of sound events from time differences of arrival and directions of arrival observed in all ARIRs for the direct sound, and in a triplet of neighboring ARIRs for early reflections. For early sound events after the direct sound, the proposed approach detects and matches peaks in the ARIR triplet that are assumed to belong to the same sound event. The resulting extrapolation is described and how interpolation artifacts are avoided by time-aligning the ARIR segments of matched sound events. In Section 5 we combine both extrapolation techniques to propose a variable-perspective ARIR rendering that is based on (i) position-dependent interpolation of jointly localized, matched ARIR peaks, and (ii) separately extrapolated, residual ARIRs; for both, rendering recombines measured ARIR segments of a triplet of perspectives (cf. Fig. 1). To support real-time operation, an offline-interpolated, fine-meshed ARIR grid is proposed to simplify the interpolation. Finally, different configurations of the proposed interpolation are evaluated in a listening experiment using both measured and simulated first-order ARIR data sets, in Section 6.

## 2 Proposed perspective interpolation

The proposed interpolation consists of the perspective extrapolation of measured or simulated Ambisonic room impulse responses (ARIRs) to the desired listener perspective and their linear interpolation within a constellation of a triplet around the listener, as shown in Figure 1.

### 2.1 ARIR-triplet interpolation

For interpolation, we propose to linearly superimpose the ARIR triplet around the listening position with weights depending on the variable position. As shown in [11, 35], a purely distance-weighted linear interpolation yields a fair directional reproduction of a recorded sound field. Therefore, we initially introduce distance weights $g_i(\boldsymbol{x}_\mathrm{d})$ that specify the contribution of each ARIR of the triplet $\boldsymbol{h}_i(t)$, $i \in \{1, 2, 3\}$ to the interpolated result by emphasizing close ARIRs and attenuating distant ones. For a horizontal, equidistant square ARIR grid with $z_i = 0$, it can be defined by,

$$g_i(\boldsymbol{x}_\mathrm{d}) = G \cdot \cos^2\left(\frac{\pi}{2r}(x_\mathrm{d} - x_i)\right) \cos^2\left(\frac{\pi}{2r}(y_\mathrm{d} - y_i)\right), \quad (1)$$

with the grid positions $\boldsymbol{x}_i = [x_i,\ y_i,\ z_i]^T$ and listener perspective $\boldsymbol{x}_\mathrm{d} = [x_\mathrm{d},\ y_\mathrm{d},\ z_\mathrm{d}]^T$. $r$ is the grid spacing of neighboring ARIRs and $G$ is chosen so that $\sum_{i=1}^{3} g_i(\boldsymbol{x}_\mathrm{d}) = 1$. The according interpolation of ARIR triplets would yield,

$$\boldsymbol{d}(t) = \sum_{i=1}^{3} g_i(\boldsymbol{x}_\mathrm{d})\,\boldsymbol{h}_i(t), \quad (2)$$

where $\boldsymbol{d}(t)$ denotes the interpolated ARIR at the desired listener perspective $\boldsymbol{x}_\mathrm{d}$.

However, the purely distance-weighted, linear interpolation alone can cause problems such as strong comb filtering artifacts due to the temporally and directionally misaligned superposition of the direct sound peaks or prominent early reflection peaks. Hereby, linear interpolation can smear such peaks either temporally or directionally, causing either an increased number of apparent peaks that are too low in amplitude or which may lack energy in the higher-order channels. Most often, this causes perceivable fluctuations of sound coloration or room impressions (distance, width) when moving through the virtual room, which is also reflected in the evaluation (cf. Sect. 6). Superior performance is expected when interpolation is preceded with perspective extrapolation, i.e. a prior parametric translation of the measured ARIR perspectives to the desired listener perspective.

### 2.2 ARIR extrapolation

The proposed perspective extrapolation decomposes an ARIR into short time segments, of which each one is interpreted as an instantaneous sound event, corresponding to an acoustic propagation path, e.g. a discrete reflection, in the room. Extrapolation ensures that sound events, such as peaks in the ARIRs, get time-aligned, level-aligned, and direction-aligned consistent with what should be received at the variable listening perspective. And yet, extrapolation should be as content-preserving as possible, and therefore its alignments in time, level, and direction within the ARIRs need to be done carefully. For instance, alignment needs not be processed individually for every sample of the ARIR if the temporal context can be preserved with a constant time shift of finite time segments. Directional alignment needs not destroy directional context if it can be done by rotation.

The perspective extrapolation of a single ARIR time instant or segment $\boldsymbol{h}_i(t)$ from the ARIR perspective $\boldsymbol{x}_i$ to the desired listener perspective $\boldsymbol{x}_\mathrm{d}$ is done assuming a known instantaneous sound-event position $\hat{\boldsymbol{x}}_t$ locating the ARIR time instant or segment in space (cf. Fig. 2).

#### 2.2.1 Rotation

As a first step, extrapolation applies a rotation to the sound-event direction of arrival (DOA) $\boldsymbol{\theta}_\mathrm{d}(\hat{\boldsymbol{x}}_t)$ that is consistent with what a listener should receive at the translated position. The rotation is accomplished by multiplying the ARIR time instant or segment with an $N$th-order spherical harmonics rotation matrix $\boldsymbol{R}_\mathrm{sp}^{(N)}(\hat{\boldsymbol{x}}_t)$ (6) that is determined by a Cartesian 3 × 3 rotation matrix $\boldsymbol{R}_{xyz}(\hat{\boldsymbol{x}}_t)$. The rotation aligns the observed DOA in form of the Cartesian unit vector $\boldsymbol{\theta}_i(\hat{\boldsymbol{x}}_t)$ with the DOA $\boldsymbol{\theta}_\mathrm{d}(\hat{\boldsymbol{x}}_t)$ at the target perspective, i.e. $\boldsymbol{\theta}_\mathrm{d}(\hat{\boldsymbol{x}}_t) = \boldsymbol{R}_{xyz}(\hat{\boldsymbol{x}}_t)\boldsymbol{\theta}_i(\hat{\boldsymbol{x}}_t)$, in azimuth and zenith,

$$\boldsymbol{R}_{xyz}(\hat{\boldsymbol{x}}_t) = \boldsymbol{R}_z(\varphi_\mathrm{d})\boldsymbol{R}_y(\vartheta_\mathrm{d} - \vartheta_i)\boldsymbol{R}_z(-\varphi_i), \quad (3)$$

$$\text{with} \quad \boldsymbol{R}_z(\alpha) = \begin{bmatrix} \cos(\alpha) & -\sin(\alpha) & 0 \\ \sin(\alpha) & \cos(\alpha) & 0 \\ 0 & 0 & 1 \end{bmatrix},$$

$$\boldsymbol{R}_y(\alpha) = \begin{bmatrix} \cos(\alpha) & 0 & -\sin(\alpha) \\ 0 & 1 & 0 \\ \sin(\alpha) & 0 & \cos(\alpha) \end{bmatrix},$$

where $\varphi_{i,\mathrm{d}}$ is the azimuth and $\vartheta_{i,\mathrm{d}}$ is the zenith of the corresponding DOA,

$$\boldsymbol{\theta}_i(\hat{\boldsymbol{x}}_t) = \frac{\hat{\boldsymbol{x}}_t - \boldsymbol{x}_i}{||\hat{\boldsymbol{x}}_t - \boldsymbol{x}_i||}, \quad \boldsymbol{\theta}_\mathrm{d}(\hat{\boldsymbol{x}}_t) = \frac{\hat{\boldsymbol{x}}_t - \boldsymbol{x}_\mathrm{d}}{||\hat{\boldsymbol{x}}_t - \boldsymbol{x}_\mathrm{d}||}. \quad (4)$$

The azimuth $\varphi$ and zenith $\vartheta$ of a Cartesian unit-length DOA vector $\boldsymbol{\theta} = [x,\ y,\ z]^T$ can be determined by,

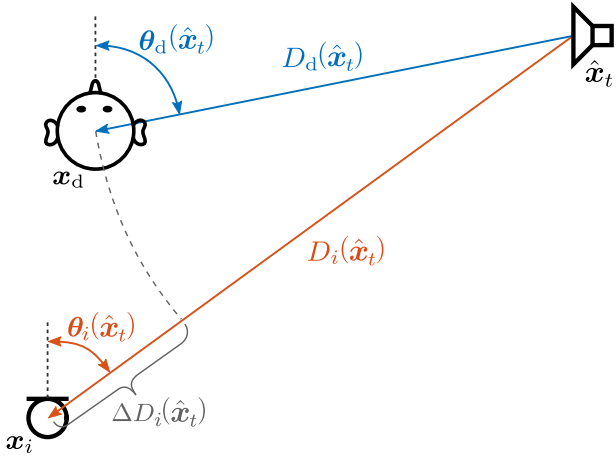

**Figure 2.** Extrapolation of an ARIR segment from its recording perspective $x_i$ to a desired listener perspective $x_d$, given its instantaneous sound-event position $\hat{x}_t$.

$$\varphi = \arctan\frac{y}{x}, \quad \vartheta = \arccos z. \tag{5}$$

The $N$th-order spherical harmonics rotation matrix,

$$\boldsymbol{R}_{\mathrm{sp}}^{(N)}(\hat{\boldsymbol{x}}_t) = R_{\mathrm{sp}}^{(N)}\{\boldsymbol{R}_{xyz}(\hat{\boldsymbol{x}}_t)\}, \tag{6}$$

is determined from (3) by recurrence relations $R_{\mathrm{sp}}^{(N)}\{\cdot\}$ according to [37]. We used the implementation contained in the *Spherical Harmonic Transform*[1] Matlab Toolbox.

### 2.2.2 Time and level adjustment

Moreover, extrapolation of an ARIR time instant or segment implies a distance shift $\Delta D_i(\hat{\boldsymbol{x}}_t)$ that physically corresponds with a shift in level (8), according to the $\frac{1}{D}$ distance law, and in time (9), according to the acoustic flight time. Altogether, the corresponding ARIR segment extrapolation becomes,

$$\vec{\boldsymbol{h}}_i(t) = g_{D,i}(\hat{\boldsymbol{x}}_t)\,\boldsymbol{R}_{\mathrm{sp}}^{(N)}(\hat{\boldsymbol{x}}_t)\,\boldsymbol{h}_i(t + \Delta t_i(\hat{\boldsymbol{x}}_t)), \tag{7}$$

$$\text{with} \quad g_{D,i}(\hat{\boldsymbol{x}}_t) = \frac{D_i(\hat{\boldsymbol{x}}_t)}{D_d(\hat{\boldsymbol{x}}_t)}, \tag{8}$$

$$\Delta t_i(\hat{\boldsymbol{x}}_t) = \frac{\Delta D_i(\hat{\boldsymbol{x}}_t)}{c} = \frac{D_i(\hat{\boldsymbol{x}}_t) - D_d(\hat{\boldsymbol{x}}_t)}{c}, \tag{9}$$

with the distances $D_i(\hat{\boldsymbol{x}}_t) = ||\hat{\boldsymbol{x}}_t - \boldsymbol{x}_i||$ and $D_d(\hat{\boldsymbol{x}}_t) = ||\hat{\boldsymbol{x}}_t - \boldsymbol{x}_d||$. The result $\vec{\boldsymbol{h}}_i(t)$ denotes the extrapolated ARIR time instant or segment that is consistently displaying the instantaneous sound-event position $\hat{\boldsymbol{x}}_t$ observed at the new perspective $\boldsymbol{x}_d$.

As first-order multi-perspective ARIR measurements are still easier to take, more easily available, or less costly because they can be taken with first-order tetrahedral microphone arrays, we propose their directional enhancement in the upcoming section, as modular tool to improve the directional resolution of first-order ARIRs to higher order. Higher-order ARIRs are not only superior in terms of their directional definition [38], we also assume their interpolation to be safer in avoiding audible interference artifacts.

---

[1] https://github.com/polarch/Spherical-Harmonic-Transform

### 2.3 Directional enhancement of first-order ARIRs

Directional enhancement is a means to improve the perceived spaciousness of first-order ARIRs, i.e. *B*-format RIRs. Merimaa and Pulkki introduced the spatial impulse response rendering (SIRR) [30], which assumes the existence of numerous time-varying narrow-band sources within an otherwise isotropic, diffuse sound field. They first estimate the diffuseness and DOA of each time-frequency bin to map non-diffuse RIR content to a loudspeaker array according to the estimated DOAs using vector based amplitude panning (VBAP). The diffuse RIR part is decorrelated and mixed to all loudspeakers. A recent work [39] introduced a SIRR approach for higher-order input and moreover investigated the perceived effect of different SIRR configurations in comparison to the first-order spatial decomposition method (SDM).

The SDM proposed by Tervo et al. [31] is more simplistic and does not differentiate between diffuse and non-diffuse RIR content and is designed in time domain only. It assumes that a single time-varying direction as carrier of the sequence of broadband sound events in the RIR is sufficient to model the directionally incoming waves. While this assumption is mostly true for the early, sparse RIR part, it does not hold for the diffuse reverberation at later times, which is characterized by multiple coincident reflections from several directions. The authors of SDM suggested to map SDM-encoded RIRs to a loudspeaker array via VBAP or to the nearest loudspeaker.

Instead of mapping first-order ARIRs to a specific loudspeaker array, recent publications propose to directively re-encode RIR sound events in higher-order Ambisonics. The SDM-based approach by Zaunschirm et al. [32] is called Ambisonic spatial decomposition method (ASDM). Similar as the earlier works, it uses a Cartesian direction-of-arrival (DOA) vector $\boldsymbol{\theta}(t)$ estimated from the smoothed pseudo intensity vector $\tilde{\boldsymbol{I}}(t)$ of a band-limited, first-order ARIR,

$$\boldsymbol{\theta}(t) = \frac{\tilde{\boldsymbol{I}}(t)}{||\tilde{\boldsymbol{I}}(t)||}, \quad \text{with} \quad \tilde{\boldsymbol{I}}(t) = F_{\mathrm{av}}\left\{W_{\mathrm{BP}}(t)\begin{bmatrix}X_{\mathrm{BP}}(t)\\Y_{\mathrm{BP}}(t)\\Z_{\mathrm{BP}}(t)\}\end{bmatrix}\right\}, \tag{10}$$

where $W(t)$ is the zeroth-order omnidirectional ARIR channel of the first-order ARIR $\boldsymbol{h}(t)$ and $\{X, Y, Z\}(t)$ are the first-order directional ARIR channels pointing to $x$, $y$, and $z$. The subscript $\{\cdot\}_{\mathrm{BP}}$ denotes zero-phase bandpass filtering between 200 Hz and 3 kHz and $F_{\mathrm{av}}\{\cdot\}$ is a zero-phase averaging filter over 10 samples at sampling rate $f_s = 44.1$ kHz, cf. [40]. This DOA estimation is also applicable to higher-order ARIRs by neglecting the channels of the order $n \geq 2$.

Subsequently, ASDM encodes the omnidirectional RIR $W(t)$ by N3D-normalized, real-valued spherical harmonics

$Y_n^m(\boldsymbol{\theta})$ of order $n$ and degree $m$, evaluated at the directions $\boldsymbol{\theta}(t)$,

$$\tilde{h}_{nm}(t) = Y_n^m[\boldsymbol{\theta}(t)]W(t), \quad n \in \{0, 1, \ldots, N\}, \quad m \in \{-n, \ldots, n\}, \tag{11}$$

where, e.g., $N = 5$ denotes the increased Ambisonic order. To revoke spectral whitening of the reverberation induced by fast DOA fluctuations whose higher-order encoding yields amplitude modulation, a spectral correction of the temporal envelope is introduced in [32], including a source-code example for MATLAB.

As stated above, the fundamental SDM assumption of only one single broad-band sound event per time instant is usually violated when reverberation gets diffuse. Therefore, we extend the ASDM directional enhancement by a decorrelation of the late ARIR part for $t > 100$ ms as described in ([38], Ch. 5) with the modulation parameters $\tau = 5$ ms and $\hat{\phi} = 50°$,

$$h_{nm}(t) = \sum_{q=0}^{5} J_{|q|}(m\hat{\phi}) \cdot \left[\cos\left(\frac{\pi}{2}|q|\right) - \text{sign}(m)\sin\left(\frac{\pi}{2}|q|\right)\right] \\ \times \tilde{h}_{s,nm}(t - q\tau), \tag{12}$$

where $J_{|q|}(\cdot)$ is the order-$|q|$ Bessel function of the first kind and $\tilde{h}_{s,nm}(t)$ are the Ambisonic channels of the spectrally corrected ARIR.

We propose the single-perspective resolution enhancement as initial step of the variable-perspective ARIR interpolation when using first-order ARIR grids. Hereafter, $\boldsymbol{h}_i(t)$ thus denotes higher-order ARIRs whose directional resolution was either enhanced from first order, which were directly captured with higher-order microphone arrays, or simulated.

## 3 Single-perspective ARIR extrapolation

As a first approach, one can think of an independent extrapolation of every ARIR, disregarding the context of neighboring ARIRs. As proposed in Section 2.2, such an independent extrapolation could be based on the estimation of an instantaneous sound-event position.

### 3.1 Instantaneous sound events in single ARIR

We propose to estimate the instantaneous sound-event position for each sample of the $i$th ARIR $\boldsymbol{h}_i(t)$ using its DOA $\boldsymbol{\theta}_i(t)$ (10) and estimated time of arrival (TOA) $t$ (note that the ARIR set should be compensated for a possible measurement delay or pre-delay truncation, which is estimated in Section 4.3),

$$\hat{\boldsymbol{x}}_t = \boldsymbol{x}_i + c \cdot t \cdot \boldsymbol{\theta}_i(t). \tag{13}$$

Each ARIR sample can be interpreted as single-sample sound event that is emitted from the according instantaneous sound-event position $\hat{\boldsymbol{x}}_t$ at time $t = 0$. Observing these spatial ARIR samples from an extrapolated listener perspective $\boldsymbol{x}_d$ would yield a sample-wise parallactic time shift, distance shift and rotation as described in Section 2.2 (7).

Figure 3a shows an exemplary trajectory of instantaneous sound event positions (gray) containing three ARIR segments of equal duration $t_1 \leq t \leq t_2$ (blue), $t_3 \leq t \leq t_4$ (red) and $t_5 \leq t \leq t_6$ (green) on straight-line trajectories. With regard to the recording perspective, the corresponding TOAs are proportional to the radial coordinate (top axis) in Figure 3a. Whereas the direction and level of each spatial ARIR sample can be directly adapted to the extrapolated listener perspective $\boldsymbol{x}_d$ according to (8) and (6), a parallactic temporal resampling per sample (9) would destroy the temporal ARIR context and yield coloration by time-scale distortion of ARIR segments, such as time-reversal (blue), temporal expansion (red) or compression (green), as displayed by the radial coordinate with regard to the listener (bottom axis) in Figure 3a. This temporal and spectral distortion may cause a distinct loss of sound quality.

### 3.2 Extrapolation restoring the temporal context

To prevent the time-scale warping, we introduce a quantized time-shift map $\Delta \bar{t}_i(\hat{\boldsymbol{x}}_t)$ constraining temporal resampling to the original sampling rate within variable-length ARIR segments. To this end, the captured ARIR is split into short-time segments, whose time shifts $\Delta t_i(\hat{\boldsymbol{x}}_t)$ (9) are quantized within each segment by its median value. The accordingly duration-preserving TOAs are displayed with regard to the listener perspective in Figure 3b along the radial coordinate (bottom axis). For most accurate results in the early ARIR, segmentation should consider detecting ARIR peaks and defining preferably long non-overlapping segments containing single distinct sound events. As we propose a more elaborate, separate extrapolation of prominent ARIR peaks that can be localized in an ARIR triplet in Section 4, the variable-length segmentation employed here is more simplistic and divides the ARIR into segments between jumps in the parallactic time shift $\Delta t_i(\hat{\boldsymbol{x}}_t)$, with short cross fades in between. Segment boundaries are accordingly defined at the extrema of $\frac{d}{dt}\Delta t_i(\hat{\boldsymbol{x}}_t)$. We implemented the extrema detection by a window of length $L$ around $t$ running over $|\Delta t_i(\hat{\boldsymbol{x}}_t)|$ starting from $t = 0$. For each time instant $t$, the global maximum within the sliding window is defined as preliminary segment boundary, whereas remaining local maxima within the window are rejected. If the preliminary segment boundary is globally maximal in all the sliding windows it is contained in, it becomes a segment boundary. For implementation, we chose $L = 16$ samples at $f_s = 44.1$ kHz, which ensures variable segment lengths of $L$ samples or more, and we employed $\frac{L}{4}$-samples $\cos^2$ cross fades between the segments. An exemplary parallactic time-shift map $\Delta t_i(\hat{\boldsymbol{x}}_t)$ (gray) and its quantized version $\Delta \bar{t}_i(\hat{\boldsymbol{x}}_t)$ (blue dotted) are shown in Figure 4 for a measured ARIR.

Extrapolation thus initially applies the computation of a rotated (10) and level-adapted (12) ARIR,

$$\tilde{\boldsymbol{h}}_i(t) = g_{D,i}(\hat{\boldsymbol{x}}_t)\boldsymbol{R}_{\text{sp}}^{(N)}(\hat{\boldsymbol{x}}_t)\boldsymbol{h}_i(t), \tag{14}$$

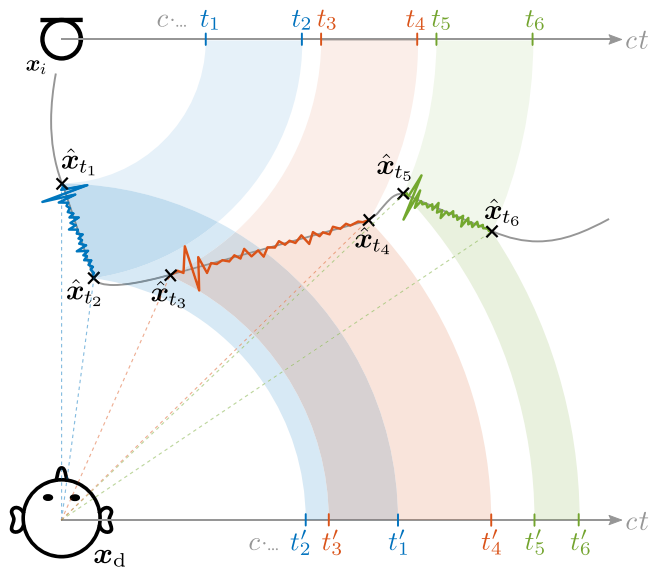

(a) Instantaneous sound-event trajectory (gray) yields level, direction, and preliminary time changes required per sample to extrapolate for a parallactically shifted listener. Problematic time mapping yields time-reversed (blue), stretched (red), or squeezed (green) ARIR segment durations along the time scale.

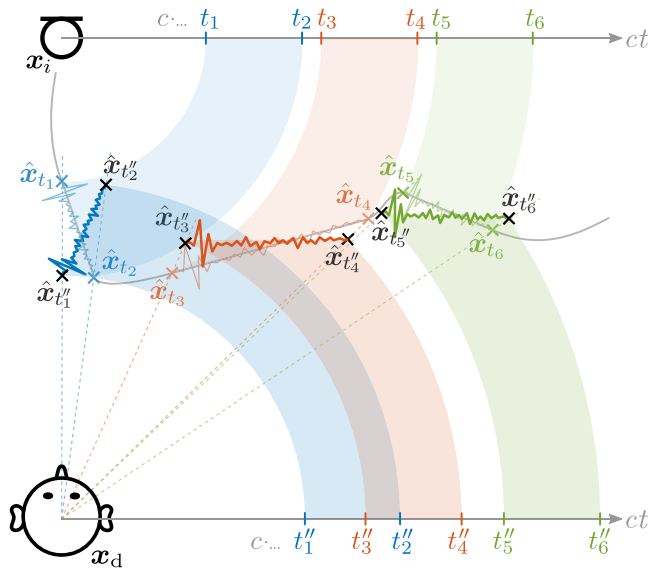

(b) Improved temporal mapping employs median quantized time shifts to map the trajectory segments $\hat{\boldsymbol{x}}_{t_1''} \to \hat{\boldsymbol{x}}_{t_2''}$, $\hat{\boldsymbol{x}}_{t_3''} \to \hat{\boldsymbol{x}}_{t_4''}$, $\hat{\boldsymbol{x}}_{t_5''} \to \hat{\boldsymbol{x}}_{t_6''}$ at their original sampling speeds while the directions of arrival are preserved.

**Figure 3.** Schematic illustration of single-perspective extrapolation by instantaneous sound-event trajectory (gray) according to (19) with three straight-line segments $\hat{\boldsymbol{x}}_{t_1} \to \hat{\boldsymbol{x}}_{t_2}$, $\hat{\boldsymbol{x}}_{t_3} \to \hat{\boldsymbol{x}}_{t_4}$, $\hat{\boldsymbol{x}}_{t_5} \to \hat{\boldsymbol{x}}_{t_6}$ of equal duration in the recorded ARIR (top time line). While level and direction re-mapping according to (a) are acceptable, temporal re-mapping (bottom time line) improves by resampling at original sampling speed, with segment time shifts quantized as their median time shift (b) (bottom time line).

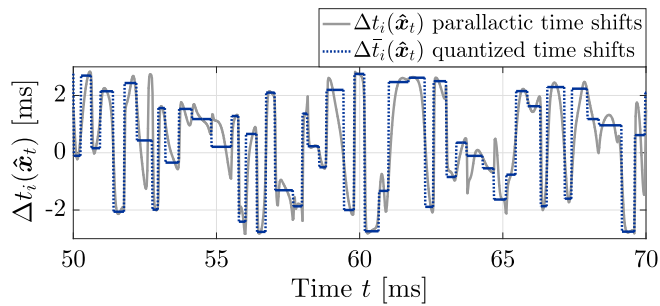

**Figure 4.** Time shifts due to parallactic resampling of the instantaneous sound-event position from an extrapolated listener perspective at a displacement of $\|\boldsymbol{x}_\mathrm{d} - \boldsymbol{x}_\mathrm{i}\| = 1$ m and segmentation with median-quantized time shifts.

and subsequently incorporates the median-quantized time shifts $\Delta \bar{t}_i(\hat{\boldsymbol{x}}_t)$ of the resulting ARIR. Finally, the windowed and time-shifted segments are superimposed to an extrapolated ARIR $\vec{h}_i(t)$.

While the assumptions employed in the perspective extrapolation approach should hold for early ARIR sound events that are most often temporally and directionally distinct, they would likely be violated in later ARIR parts. Therefore, rather than re-encoding at single directions, rotation is considered to more robustly preserve non-unique directional content. Moreover, studies on mixing time [41, 42] indicate that sufficiently late parts can be exchanged without audible effects, as their spectral properties are quite similar after a change in perspective. We restrict the perspective extrapolation to the first 100 ms as this adjustable limit worked well for both scenarios evaluated in Section 6.

## 4 Multi-perspective ARIR extrapolation

The extrapolation in Section 3 can only deliver an adjustment of delay, gain, and direction of an individual ARIR perspective ($P = 1$) to the listening position, devoid of possible context that could be retrieved from multiple ARIR perspectives.

Whenever clear constellations of arrival times are observed in ARIRs captured at neighboring positions, a more systematic, joint sound-event localization becomes possible. We assume clear constellations to apply to high-energy sound events in the early, sparse ARIRs, such as the direct-sound peak or early reflections, which would provide an accurate estimation of arrival times. Consequently, a corresponding extrapolation from jointly localized temporal peaks as sound events could avoid the sample-based rotations and time shifts from Section 3.2, and it could hereby preserve longer temporal and directional contexts within time segments around high-energy ARIR peaks. Moreover, localization based on arrival times is assumed to offer a substantially higher localization accuracy compared to a single-DOA-based localization, because of the large spacing between neighboring perspectives.

This section first introduces the detection and fundamental parameter estimation of ARIR peaks. While the targeted variable-perspective ARIR rendering is applicable to ARIRs of arbitrary order, the parameters are estimated from components of the zeroth and the first order. Secondly, the proposed approaches for joint sound-event localization are introduced, which in particular include a global localization of the direct-sound source and a triplet-based sound-event localization of early peaks, that could, e.g., relate to image source positions of early reflections, based on the observed TDOAs and DOAs. Finally, the extrapolation of ARIR segments around jointly localized sound events is described.

## 4.1 TOA, DOA and magnitude of ARIR peaks

To detect meaningful time instants of sound events, we propose an ARIR peak detection based on a short-time magnitude of the directional content, represented by the envelope of both the omnidirectional and the first-order directional ARIR channels. It can be computed from the averaged magnitude of the pseudo intensity vector $\boldsymbol{I}(t)$,

$$\bar{a}(t) = \sqrt{F_\mathrm{H}\{||\boldsymbol{I}(t)||\}}, \quad \text{with} \quad \boldsymbol{I}(t) = W(t)\begin{bmatrix} X(t) \\ Y(t) \\ Z(t) \end{bmatrix}, \quad (15)$$

without band limitation, and here $F_\mathrm{H}\{\cdot\}$ denotes a Hamming-windowed moving-average filter over 0.5 ms. Each maximum of $\bar{a}(t)$ that overshoots a predefined prominence threshold is defined as peak with time of arrival (TOA) $T$. The threshold is best chosen depending on the reverberation of the room, trading off the number of distinct peaks detected and their reliability by disposing low-energy peaks. From the time-varying DOAs of (10), a static DOA $\boldsymbol{\theta}(T)$ can be assigned to each time segment containing a detected peak.

In summary, we get a characteristic TOA $T$, minus a general offset that is yet unknown (cf. Sect. 4.3), a magnitude $\bar{a}(T)$, and the DOA $\boldsymbol{\theta}(T)$ of each prominent peak detected in the early ARIR $\boldsymbol{h}(t)$.

## 4.2 TDOA-based sound-event localization

DOA-based source localization is known from approaches using Ambisonic recordings from multiple perspectives [13, 14, 43]. However, a purely DOA-based localization of sound events from ARIR peaks can be inferior in precision. An arrival-time-based source localization offers a reasonable alternative of potentially higher accuracy. Since measured RIRs can contain an unknown system delay, or more often than not a common pre-delay is removed, their time differences of arrival (TDOAs) are considered as information available from a grid of measured multi-perspective RIRs or ARIRs. We propose a predominantly TDOA-based localization to achieve accuracy and robustness to measurement uncertainties.

### 4.2.1 Least-squares localization

A detailed overview over existing passive source localization approaches is given in [44]. We suggest a localization based on the spherical least-square (LS) error (16) introduced by Schau et al. in [45]. For mathematical simplification, the coordinate system is initially shifted so that an arbitrary ARIR of the grid, defined as the first ARIR, is located in the origin, i.e. $\boldsymbol{x}_1 = \vec{\boldsymbol{0}}$. The 3-dimensional spherical LS error function is defined by,

$$\boldsymbol{e}_\mathrm{sp}(\boldsymbol{x}_t) = \boldsymbol{S}\boldsymbol{x}_t + r_t\hat{\boldsymbol{d}} - \boldsymbol{b}, \quad \text{with} \quad \boldsymbol{S} = \begin{bmatrix} x_2 & x_3 & \ldots & x_P \\ y_2 & y_3 & \ldots & y_P \\ z_2 & z_3 & \ldots & z_P \end{bmatrix}^\mathrm{T},$$

$$\boldsymbol{x}_t = \begin{bmatrix} x_t \\ y_t \\ z_t \end{bmatrix}, \quad \hat{\boldsymbol{d}} = \begin{bmatrix} \hat{d}_{21} \\ \hat{d}_{31} \\ \vdots \\ \hat{d}_{P1} \end{bmatrix}, \quad \boldsymbol{b} = \frac{1}{2}\begin{bmatrix} r_2^2 - \hat{d}_{21}^2 \\ r_3^2 - \hat{d}_{31}^2 \\ \vdots \\ r_P^2 - \hat{d}_{P1}^2 \end{bmatrix}, \quad (16)$$

where $r_t = ||\boldsymbol{x}_t - \boldsymbol{x}_1|| = ||\boldsymbol{x}_t||$ is the unknown distance of the unknown sound-event position $\boldsymbol{x}_t$ to the first ARIR perspective at $\boldsymbol{x}_1 = \vec{\boldsymbol{0}}$. $r_i = ||\boldsymbol{x}_i - \boldsymbol{x}_1|| = ||\boldsymbol{x}_i||$ is the known setup distance between the $i$th and the first ARIR perspective for $i \in \{2, \ldots, P\}$, where $P$ denotes the total number of ARIR perspectives included in localization. Furthermore,

$$\hat{d}_{i1} = c \cdot \hat{\tau}_{i1} = c \cdot (T_i - T_1), \quad (17)$$

are the observed TDOA-dependent range differences between the $i$th and the first ARIR according to the peak TOAs $T_i$ and $T_1$, where $c$ is the speed of sound.

Equation (16) is designed for 3-dimensionally distributed receiver arrays. The 3D spherical LS error is equivalent to a 2D spherical LS error for a purely horizontal ARIR grid as the last column in $\boldsymbol{S}$ with $z_i = 0, \forall i$ vanishes and suppresses $z_t$, leaving only a dependency of $r_t$ on the height $z_t$ of the sound event,

$$\boldsymbol{e}_\mathrm{sp,2D}(\boldsymbol{x}_t) = \boldsymbol{S}_\mathrm{2D}\boldsymbol{x}_{t,\mathrm{2D}} + r_t\hat{\boldsymbol{d}} - \boldsymbol{b}, \quad \text{with}$$

$$\boldsymbol{S}_\mathrm{2D} = \begin{bmatrix} x_2 & x_3 & \cdots & x_P \\ y_2 & y_3 & \cdots & y_P \end{bmatrix}, \quad x_{t,\mathrm{2D}} = \begin{bmatrix} x_t \\ y_t \end{bmatrix}. \quad (18)$$

Accordingly, the LS cost function yields

$$J_\mathrm{sp,2D}(\boldsymbol{x}_t) = \boldsymbol{e}_\mathrm{sp,2D}^\mathrm{T}(\boldsymbol{x}_t)\,\boldsymbol{e}_\mathrm{sp,2D}(\boldsymbol{x}_t). \quad (19)$$

It vanishes at its optimum for $P = 3$ and exhibits non-zero minima for $P > 3$, when assuming the presence of typical measurement uncertainties. In this case, sound-event localization is done by minimizing,

$$\hat{\boldsymbol{x}}_{t,\mathrm{LS}} = \arg\min_{\boldsymbol{x}_t} J_\mathrm{sp,2D}(\boldsymbol{x}_t). \quad (20)$$

### 4.2.2 Non-uniqueness in height $z_t$

As the 2D spherical LS cost function only contains a dependency of $r_t$ on $z_t$, the sign of $z_t$ is not determined and $J_\mathrm{sp,2D}([x_t, y_t, z_t]^T) = J_\mathrm{sp,2D}([x_t, y_t, -z_t]^T)$. For $P > 3$, we get two minimum-LS sound-event locations $\hat{\boldsymbol{x}}_{t,\mathrm{LS}}$ at $\pm z_{t,\mathrm{LS}}$, in general. For $P = 3$, the LS cost function vanishes for any pre-selected $z_t$, yielding infinitely many sound-event locations on a vertically symmetric hyperbola (cf. Fig. 5). In either case, $P = 3$ or $P > 3$, the 2D LS criterion alone is non-unique. To solve this ambiguity, we extend the LS cost function by an angular error function (21). Specifically, the DOAs of the resulting ambiguous sound-event candidates $\boldsymbol{\theta}_{x,i}(\boldsymbol{x}_t)$ are projected on the observed sound-event DOAs $\boldsymbol{\theta}_i(T_i)$, $\forall i$,

$$J_{\mathrm{ang}}(\boldsymbol{x}_t) = \sum_i \left[ 1 - \boldsymbol{\theta}_i^T(T_i)\boldsymbol{\theta}_{x,i}(\boldsymbol{x}_t) \right], \quad \text{with}$$
$$\boldsymbol{\theta}_{x,i}(\boldsymbol{x}_t) = \frac{\boldsymbol{x}_t - \boldsymbol{x}_i}{||\boldsymbol{x}_t - \boldsymbol{x}_i||}. \quad (21)$$

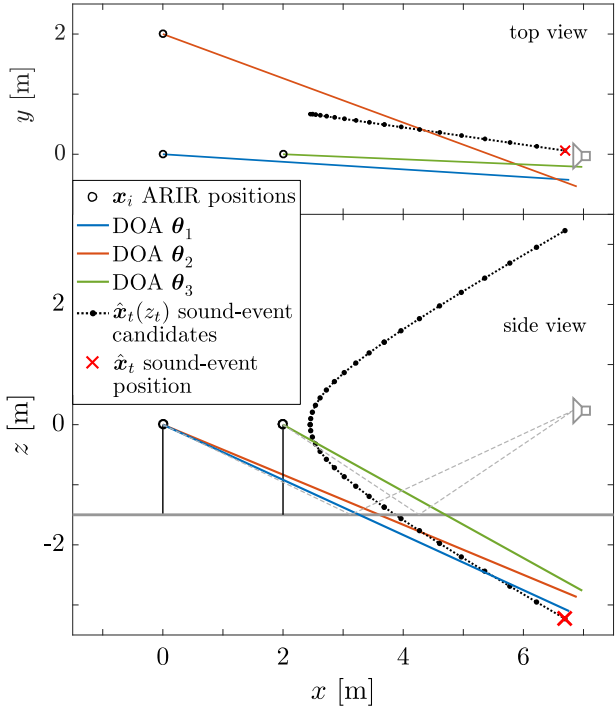

**Figure 5.** TDOA- and DOA-based sound-event localization of the first early reflection (first-order floor reflection) in an ARIR triplet. The cross marks the detected sound-event position, i.e. the point on the TDOA-localized hyperbola with minimum angular deviation to the estimated DOAs. The direct-sound source is displayed as gray loudspeaker.

The unique sound-event location is found as minimum of,

$$\hat{\boldsymbol{x}}_t = \arg\min_{\hat{\boldsymbol{x}}_{t,\mathrm{LS}}} J_{\mathrm{ang}}(\hat{\boldsymbol{x}}_{t,\mathrm{LS}}). \quad (22)$$

### 4.2.3 Ambiguity of peak combinations

While the first peak in every ARIR typically belongs to a single direct-sound event, assuming the absence of occlusion, the grouping of later arrival time constellations into distinct sound events is not as trivial. After the first peak, possible ARIR peak constellations can overlap in time, yielding a combinatorial ambiguity. The number of possible peak combinations rises drastically when many ARIRs are included. This is because the number of peak constellations to be matched grows exponentially with the flight time range within which physical peak combinations are searched. Therefore, a global matching of early ARIR peaks including a global localization of corresponding sound-event positions is impractical.

We propose global localization only for the first peak in every ARIR ($P > 3$) to get a stable direct-sound event. For the most prominent early ARIR peaks after the direct sound, we propose to use sound-event localization and matching within triplets of neighboring ARIR perspectives ($P = 3$) (cf. Sect. 4.5), taking into account salient peak features (cf. Sect. 4.1). Remaining, less distinct early ARIR peaks, especially the later peaks present in the early ARIR part, may be smeared making its correct detection and matching difficult. But by their low prominence, these peaks are also assumed to be less critical so that single-ARIR sound-event localization ($P = 1$) should work accurately enough for them.

### 4.3 Global direct sound localization

As the direct-sound event is typically predominant and specifies the perceived direction due to the precedence effect, its extrapolation imposes the highest consistency and smoothness requirement when rendering for a listener that moves through different local ARIR triplets. To ensure a perfectly stable trajectory of the direct-sound event, we suggest a global direct-sound localization using all ARIR perspectives that receive the direct-sound event, if the direct-sound location is not known from the ARIR measurements. The TDOAs between direct-sound peaks in multiple perspectives can be estimated at high accuracy and enable robust localization. This can be accomplished, e.g., using the linear correction least-squares (LCLS) estimator [46] to minimize the 2D spherical LS cost function in Equation (13) by a constrained optimization. The optimization procedure is presented in detail in [44, 46]. As stated in Section 4.2, the global TDOA localization with more than three ARIR perspectives yields two possible minimum-LS direct-sound locations that are symmetric with regard to the horizontal ARIR plane. A unique direct-sound event location $\hat{\boldsymbol{x}}_s$ is selected by minimizing the angular error function (21).

Apart from a stable direct-sound location, the global direct-sound localization allows to estimate and compensate for a uniform system delay (as also used in Sect. 3). Moreover, microphone positioning errors of ARIR measurements, such as unintended rotations, can be detected and readjusted by an ARIR rotation according to the estimated direct-sound DOA.

### 4.4 ARIR-triplet sound-event localization

In contrast to the direct-sound event, small location fluctuations can be assumed to stay inaudible when the listener moves through different local ARIR triplets. Therefore, a TDOA-based localization in ARIR triplets offers sufficient accuracy at minimal combinatorial complexity when matching possible peak combinations.

We propose to use the spherical intersection (SX) estimator [45] as TDOA-based sound event localizer, for which the LS error function (18) of a perspective triplet ($P = 3$) is zeroed. It is computationally less complex than the LCLS approach and moreover offers a closed-form solution. As the SX estimator is originally designed for three-dimensional receiver arrays and $P \geq 3$, we adapt the formalism to purely horizontal receiver triplets.

Zeroing the 2D spherical LS error (36) for an ARIR triplet with $i \in \{1, 2, 3\}$ and solving for $\boldsymbol{x}_{t,\text{2D}}$ yields,

$$\boldsymbol{x}_{t,\text{2D}} = \boldsymbol{S}_{\text{2D}}^{-1}(\boldsymbol{b} - r_t \hat{\boldsymbol{d}}). \quad (23)$$

This is equal to a straight line equation, which implies that $\boldsymbol{x}_{t,\text{2D}}$ is a function of the unknown sound-event distance $r_t$. The relation to the 3D coordinates lies in,

$$r_t^2 \stackrel{!}{=} ||\boldsymbol{x}_t||^2 = \boldsymbol{x}_t^\mathrm{T} \boldsymbol{x}_t = \boldsymbol{x}_{t,\text{2D}}^\mathrm{T} \boldsymbol{x}_{t,\text{2D}} + z_t^2$$
$$= [\boldsymbol{S}_{\text{2D}}^{-1}(\boldsymbol{b} - r_t \hat{\boldsymbol{d}})]^\mathrm{T} [\boldsymbol{S}_{\text{2D}}^{-1}(\boldsymbol{b} - r_t \hat{\boldsymbol{d}})] + z_t^2, \quad (24)$$

which is transformed to the quadratic equation,

$$\alpha r_t^2 + \beta r_t + \gamma = 0, \quad \text{with} \quad \alpha = 1 - \hat{\boldsymbol{d}}^\mathrm{T} (\boldsymbol{S}_{\text{2D}}^{-1})^\mathrm{T} \boldsymbol{S}_{\text{2D}}^{-1} \hat{\boldsymbol{d}},$$
$$\beta = 2 \boldsymbol{b}^\mathrm{T} (\boldsymbol{S}_{\text{2D}}^{-1})^\mathrm{T} \boldsymbol{S}_{\text{2D}}^{-1} \hat{\boldsymbol{d}}, \quad \gamma = -\boldsymbol{b}^\mathrm{T} (\boldsymbol{S}_{\text{2D}}^{-1})^\mathrm{T} \boldsymbol{S}_{\text{2D}}^{-1} \boldsymbol{b} - z_t^2. \quad (25)$$

Equation (25) indicates that $r_t$ relates to the absolute value of the unknown sound-event height $|z_t|$. Source localization with TDOA triplets turns out to yield an arbitrary source position on a vertical hyperbola that is symmetric with regard to the horizontal plane of the ARIR triplet (cf. Fig. 5). The sound-event position candidates are found as positive and real solution $\hat{r}_t(z_t) \in \mathbb{R}^+$ of (25) for any preselected $z_t$, which completes (23) to,

$$\hat{\boldsymbol{x}}_t(z_t) = \begin{bmatrix} \boldsymbol{S}_{\text{2D}}^{-1}(\boldsymbol{b} - \hat{r}_t(z_t) \hat{\boldsymbol{d}}) \\ z_t \end{bmatrix}. \quad (26)$$

A unique sound-event position is found by the above-mentioned angular error function (21). Specifically, we search the sound-event candidate that minimizes the angular error within a predefined range of possible $z_t$ values that is limited by the maximum time of flight $T_i$,

$$\hat{\boldsymbol{x}}_t = \arg\min_{\hat{\boldsymbol{x}}_t(z_t)} J_{\text{ang}}(\hat{\boldsymbol{x}}_t(z_t)), \quad \text{with} \quad (27)$$
$$z_t \in \{\mathbb{R} : ||\hat{\boldsymbol{x}}_t(z_t) - \boldsymbol{x}_i|| \leq c \cdot T_i\} \, \forall \, i \in \{1, 2, 3\}.$$

For the implementation of (27), we propose a grid search algorithm in steps of e.g. $\Delta z_t = 0.1$ m, yielding a sufficiently accurate result at low implementation and computation effort.

### 4.5 ARIR-triplet peak matching

The objective of the peak matching is to find peaks in a triplet of neighboring ARIRs that correspond to commonly detected sound events, for instance such stemming from the image sources of early reflections (cf. Figs. 6 and 7). To this end, we initially pre-select peak combinations within a window of feasible flight time differences observed by the ARIR triplet. From these pre-selected arrival times, matching combinations get picked, whose TDOA-based sound-event locations are geometrically consistent with the DOAs and amplitudes observed in the three ARIRs.

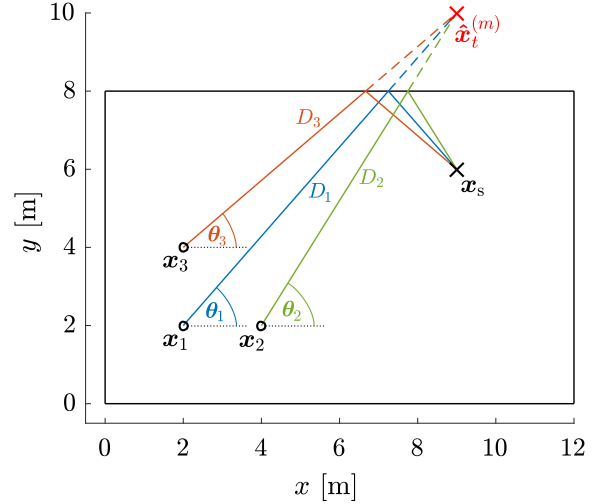

**Figure 6.** First-order wall reflection, when $\boldsymbol{x}_s$ is the direct-sound source and $\hat{\boldsymbol{x}}_t^{(m)}$ is the sound-event position of the $m$th peak match corresponding to an image source.

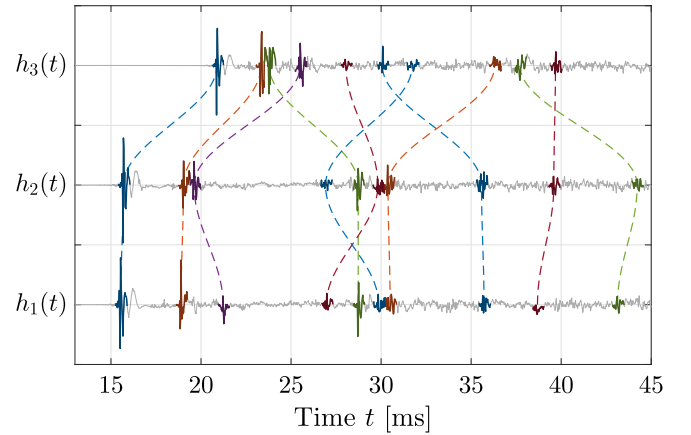

**Figure 7.** Peak matching result of the 10 greatest ARIR triplet peaks (direct-sound peak and 9 early reflections).

In detail, we propose an iterative peak matching that always starts with the greatest unmatched peak in the ARIR triplet $\boldsymbol{h}_i(t)$, $i \in \{1, 2, 3\}$ as reference peak. Hereafter, we denote this reference peak TOA by $T_A$ and the corresponding ARIR perspective by $\boldsymbol{x}_A$, $A \in \{1, 2, 3\}$. All unmatched peaks in the remaining two inferior ARIRs $B$, $C \in \{1, 2, 3\}$ can be limited to those TOAs $T_{B,C}$ whose TDOAs stay within a time window of the maximally observable flight time difference,

$$|T_A - T_{B,C}| \leq \frac{1}{c} ||\boldsymbol{x}_A - \boldsymbol{x}_{B,C}||, \quad (28)$$

as defined by the spacing of the two inferior, neighboring ARIR perspectives $\boldsymbol{x}_{B,C}$ to the one observing the reference peak, i.e. $\boldsymbol{x}_A$. Whenever any of the two inferior ARIRs exhibits more than one peak TOA candidate within this window, multiple candidates of the TOAs $T_B$ and $T_C$ could be combined with the TOA $T_A$ of the reference peak to a TDOA triplet $\{T_A, T_B, T_C\}$ for localization. The following cost function is introduced to retrieve the combination that is most consistent,

$$J(\hat{\boldsymbol{x}}_t) = J_{\mathrm{ang}}(\hat{\boldsymbol{x}}_t) + J_{\mathrm{amp}}(\hat{\boldsymbol{x}}_t). \tag{29}$$

It penalizes geometrical inconsistency with regard to directions and amplitudes observed according to the jointly localized sound-event position $\hat{\boldsymbol{x}}_t$, which is initially estimated for each pre-selected TOA-triplet using the adapted SX-based localization approach described in Section 4.4. The cost function (29) involves the angular mismatch $J_{\mathrm{ang}}(\hat{\boldsymbol{x}}_t)$ (21) between the estimated peak DOAs $\boldsymbol{\theta}_i(T_i)$ and the direction of each sound-event position $\boldsymbol{\theta}_{\boldsymbol{x},i}(\hat{\boldsymbol{x}}_t)$ from the $i$th ARIR perspective, for $i \in \{1, 2, 3\}$. Moreover, it contains a peak amplitude error (32) considering the ratio of peak amplitudes $\bar{a}_i(T_i)$ according to the estimated sound-event distance. Assuming free sound propagation, the $\frac{1}{D}$ distance law demands,

$$\bar{a}_i(T_i) \propto \frac{1}{D_i} \quad \mathrm{with} \quad D_i = ||\hat{\boldsymbol{x}}_t - \boldsymbol{x}_i||, \tag{30}$$

which can be reformulated by,

$$\varrho_i(\hat{\boldsymbol{x}}_t, T_i) = \frac{\bar{a}_i(T_i)}{||\hat{\boldsymbol{x}}_t - \boldsymbol{x}_i||} \stackrel{!}{=} \mathrm{const.} \quad \forall i \in \{1, 2, 3\}. \tag{31}$$

However, as for multiple wall reflections with unknown attenuation, higher-order reflections need not necessarily fulfill free field attenuation conditions. Hence, we suggest to relax the distance law criterion by attenuating the denominator of (31) with an exponential factor $\alpha(t)$, assuming equal peak amplitudes of later coincident reflections. We define $\alpha(t = T_{i,\mathrm{DS}}) = 1$ ($\frac{1}{D}$ distance law) for the direct-sound peak and $\alpha(t > T_{i,\mathrm{DS}} + 50\ \mathrm{ms}) \to 0$ (equal peak amplitudes) for later reflections. The peak amplitude error function measures the deviation of the resulting weighted peak amplitudes $\tilde{\varrho}_i(\hat{\boldsymbol{x}}_t, T_i)$,

$$J_{\mathrm{amp}}(\hat{\boldsymbol{x}}_t) = \frac{1}{\sqrt{3} - 1} \left( \sqrt{3} \cdot \frac{\sqrt{\sum_{i=1}^{3} \tilde{\varrho}_i(\hat{\boldsymbol{x}}_t, T_i)^2}}{\sum_{i=1}^{3} \tilde{\varrho}_i(\hat{\boldsymbol{x}}_t, T_i)} - 1 \right), \tag{32}$$

$$\mathrm{with} \quad \tilde{\varrho}_i(\hat{\boldsymbol{x}}_t, T_i) = \frac{\bar{a}_i(T_i)}{||\hat{\boldsymbol{x}}_t - \boldsymbol{x}_i||^{\alpha(T_i)}}.$$

Both, the angular error (21) and the peak amplitude error (32) are limited to values $\leq 1$.

For each reference peak, the most consistent result can thus be selected from all possible combinations that fulfill (28) by minimizing the inconsistency $J(\hat{\boldsymbol{x}}_t)$ (29). The corresponding sound-event position is localized by,

$$\hat{\boldsymbol{x}}_t^{(m)} = \arg\min_{\hat{\boldsymbol{x}}_t} J(\hat{\boldsymbol{x}}_t), \quad \forall m \geq 2, \tag{33}$$

where $m \in \{1, \ldots, M\}$ denotes the peak matching index. $m = 1$ corresponds to the matched direct-sound peaks with the globally localized source position $\hat{\boldsymbol{x}}_t^{(1)} = \hat{\boldsymbol{x}}_s$. Consequently, $m = 2$ denotes the match of the greatest early reflection corresponding to the triplet-localized sound-event position $\hat{\boldsymbol{x}}_t^{(2)}$, which most likely corresponds to an image source of a first-order reflection. The peak matching is iterated over the next unmatched reference peak, until a desired number of peak matches $M$ has been found.

### 4.6 Extrapolation of matching ARIR peak segments

This section takes up the information about matched peaks in ARIR-triplets to commonly extrapolate ARIR segments around those matched peaks to the desired listener perspective $\boldsymbol{x}_\mathrm{d}$.

Initially, ARIR segments $h_{\mathrm{p},i}^{(m)}(t)$ of equal length are cut around the peaks of each matched peak triplet $m \in \{1, \ldots, M\}$. We define each segment to start 16 samples before the estimated peak TOA $T_i$ and end at least 16 samples before the earliest successive peak, however limited to a total segment length staying below 3 ms. This allows to preserve temporal, directional and level information present in the ARIR peak segments in a preferably long-enough context. For smooth transitions between ARIR segments, the boundaries of each segment are cross-faded by an overlapping $\cos^2$ half-window of 16-samples length.

Subsequently, we shift matching ARIR peak segments consistently with regard to the current location of a variable-perspective listener position by perspective extrapolation as described in Section 2.2 (7) with $\hat{\boldsymbol{x}}_t = \hat{\boldsymbol{x}}_t^{(m)}$. The resulting extrapolated ARIR segment of the $i$th ARIR according to the $m$th peak match with sound-event position $\hat{\boldsymbol{x}}_t^{(m)}$ is denoted by $\vec{h}_{\mathrm{p},i}^{(m)}(t)$.

## 5 Variable-perspective ARIR rendering

We proposed a system for rendering ARIRs in six degrees of freedom (6DoF) employing the ARIR perspective extrapolation with information retrieved (i) from all the ARIRs for the direct-sound event, (ii) from ARIR triplets for early sound events, and (iii) from individual ARIRs for the remaining residual sound events. The system targets a variable-perspective interpolation only using the extrapolated ARIRs of the perspective triplet around the listener position, within a horizontally distributed ARIR grid of arbitrary Ambisonic order $N \geq 1$. As most ARIR grids still are first-order Ambisonic (e.g. [34]), directional enhancement of first-order ARIRs offers a meaningful improvement. For this purpose, we integrate the ASDM directional enhancement as optional component of the proposed variable-perspective ARIR rendering. Figure 8 shows the block diagram of the full ARIR interpolation system.

We assume the extrapolation of the jointly, globally localized direct-sound event to be most accurate and temporally/directionally context-preserving, followed by the ARIR-triplet-based early sound-event extrapolation, and finally the single-ARIR-based extrapolation. While the extrapolation of direct and early sound events is well-defined, we build a distinction to the single-ARIR extrapolation in terms of an ARIR residual. Residual ARIR components are defined as parts that could not be jointly localized in the triplet, or such that are diffuse.

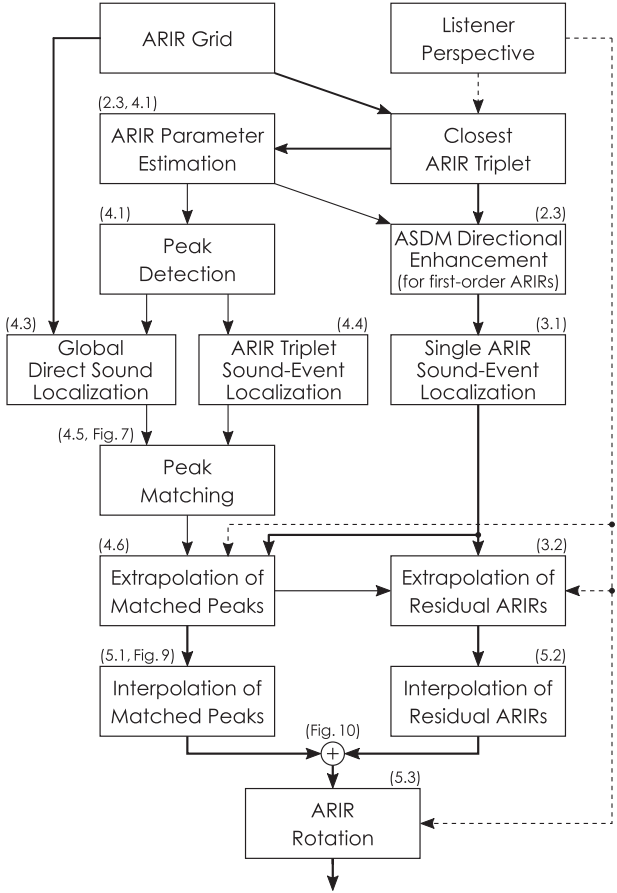

**Figure 8.** Block diagram of the proposed variable-perspective ARIR rendering using a parametric, position-dependent interpolation of a horizontally distributed ARIR triplet.

### 5.1 Interpolation of direct sound and matched peaks

Section 4 described the matching of the $M$ most distinct ARIR peak triplets and the time-aligned perspective extrapolation of corresponding ARIR peak segments to the desired listener position according to a globally-localized direct-sound source and triplet-localized reflection sound events.

While geometric time alignment of the hereby extrapolated ARIR segments $\vec{h}_{\mathrm{p},i}^{(m)}(t)$ is considered to be already rather precise in matching the two ARIRs containing the inferior peak segments to their corresponding $m$th ARIR reference peak segment, temporal misalignment in the range of a few samples can still occur. We propose a refined time alignment by shifting the inferior ARIR peak segments as to maximize their cross-correlations with their corresponding reference ARIR peak, for each matching index $m$. Hereby, a predominantly constructive superposition of ARIR peak samples can be ensured for their linear interpolation.

From the accordingly extrapolated and time-aligned peak segments of the three neighboring ARIRs, we compute an interpolated ARIR peak segment $\tilde{\boldsymbol{d}}_{\mathrm{p}}^{(m)}(t)$ for each peak match $m$ at the variable listener perspective by linear, distance-weighted (1) superposition of $\vec{h}_{\mathrm{p},i}^{(m)}(t)$,

$$\tilde{\boldsymbol{d}}_{\mathrm{p}}^{(m)}(t) = \sum_{i=1}^{3} g_i(\boldsymbol{x}_{\mathrm{d}})\, \vec{h}_{\mathrm{p},i}^{(m)}(t). \tag{34}$$

Regardless of the time alignment and its refinement, there may still be components that can only add up stochastically rather than additively. To avoid unintentional level variations between the individual, interpolated ARIR peak segments, we propose to control and correct the resulting RMS level of each interpolated peak by a gain factor,

$$\boldsymbol{d}_{\mathrm{p}}^{(m)}(t) = \sqrt{P_{\mathrm{cor}}^{(m)}(t)} \cdot \tilde{\boldsymbol{d}}_{\mathrm{p}}^{(m)}(t), \quad \text{with}$$

$$P_{\mathrm{cor}}^{(m)}(t) = \frac{\sum_{i=1}^{3}\left(g_i(\boldsymbol{x}_{\mathrm{d}}) \cdot \sum_{t} \vec{h}_{\mathrm{p},i}^{(m)}(t)^2\right)}{\sum_{t} \tilde{d}_{\mathrm{p}}^{(m)}(t)^2}, \tag{35}$$

where $\vec{h}_{\mathrm{p},i}^{(m)}(t)$ and $\tilde{d}_{\mathrm{p}}^{(m)}(t)$ are the omnidirectional, zeroth-order channels of $\vec{\boldsymbol{h}}_{\mathrm{p},i}^{(m)}(t)$ respectively $\tilde{\boldsymbol{d}}_{\mathrm{p}}^{(m)}(t)$. Finally, the ARIR $\boldsymbol{d}_{\mathrm{p}}(t)$ containing all interpolated matched peak segments is computed by,

$$\boldsymbol{d}_{\mathrm{p}}(t) = \sum_{m=1}^{M} \boldsymbol{d}_{\mathrm{p}}^{(m)}(t). \tag{36}$$

Because of the effort required, we propose to restrict the matching and interpolation procedure to a limited number of loudest early ARIR peaks within the first 50...75 ms after the direct sound, e.g., about $M = 10$ matches for a medium-sized room. This choice performed well in the scenarios evaluated in Section 6 and mostly comprises all prominent early reflections in about half the mixing time [41, 42]. It moreover prevents peak matching errors that potentially arise from lower-amplitude peaks thereafter, which gradually become less sparse. One can consider suitably enlarging this number of peak matches to more accurately present prominent early reflections for more complex, sparse reflection geometries or larger rooms. Figure 9 shows the result of the position-dependent interpolation of $M = 10$ matched peaks of an ARIR triplet with $r = 2$ m spacing (blue). Apart from the obvious gaps between the peaks, this interpolation reproduces a control RIR (gray) measured between the ARIR perspectives fairly well.

### 5.2 Interpolation of ARIR residuals

The sound events jointly localized in an ARIR triplet correspond to peak segments that were extrapolated and interpolated at high accuracy. Removing those segments

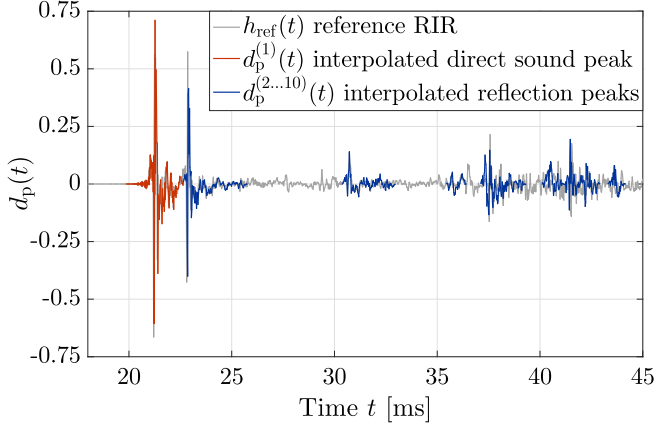

**Figure 9.** Interpolated matched peaks $d_p(t)$ according to a globally localized direct-sound event and triplet-localized reflection-sound events in comparison to a reference RIR $h_{\text{ref}}(t)$ measured at the target listener position.

from the three ARIRs of the triplet leaves three residual ARIRs for $i \in \{1, 2, 3\}$,

$$h_{r,i}(t) = h_i(t) - \sum_{m=1}^{M} h_{p,i}^{(m)}(t). \quad (37)$$

Even though the most prominent peaks in early residual ARIRs are zeroed, there may still be valuable non-diffuse parts in between, with distinct directional information, such as smaller, poorly localizable early reflections or diffraction. As these three residuals may moreover convey important parts of the room impression, we desire a technique for their extrapolation from any of the three perspectives to the desired perspective of the listener before superposition. For this purpose, the single-ARIR extrapolation approach based on the simplistic instantaneous sound-event localization described in Section 3 is applicable. After single-perspective extrapolation of each early residual ARIR of the triplet to the desired listener perspective, we interpolate the resulting residual ARIRs by linear, distance-weighted combination,

$$\tilde{d}_r(t) = \sum_{i=1}^{3} g_i(x_d) \vec{h}_{r,i}(t), \quad (38)$$

where $\vec{h}_{r,i}(t)$ are the extrapolated residual ARIRs according to Section 3.2. The spectrum of the interpolated residual $\tilde{d}_r(t)$ can depend on the particular interference as well as on remaining spectral artifacts due to the segmentation and resampling of extrapolated residuals. To compensate for this, we apply a correction of the temporal spectral envelopes, similar to the one done for ASDM directional enhancement described in [32]. Specifically, this correction restores the short-time energy of the interpolated residual ARIR (41) to match the desired, average short-time energy of the unprocessed residual ARIRs (40), in third-octave bands. This is accomplished by multiplication with the third-octave weights $w_n^b(t)$ depending on the Ambisonic order and the time instant,

$$d_{r,nm}(t) = \sum_b w_n^b(t) \cdot F_b\{\tilde{d}_{r,nm}(t)\} \quad \text{with}$$

$$w_n^b(t) = \sqrt{\frac{(2n+1) \cdot \bar{P}_{\text{ref}}^b(t)}{\sum_{m=-n}^{n} \bar{P}_{r,nm}^b(t)}}, \quad (39)$$

$$\bar{P}_{\text{ref}}^b(t) = \sum_{i=1}^{3} g_i(x_d) F_{\text{av}}\{F_b\{h_{r,i}(t)\}^2\}, \quad (40)$$

$$\bar{P}_{r,nm}^b(t) = F_{\text{av}}\{F_b\{\tilde{d}_{r,nm}(t)\}^2\}. \quad (41)$$

Here, $h_{r,i}(t)$ are the omnidirectional, zeroth-order channels of the residual ARIRs $h_{r,i}(t)$ and $\tilde{d}_{r,nm}(t)$ is the order $n$ and degree $m$ channel of the interpolated residual $\tilde{d}_r(t)$. $F_b\{\cdot\}$ is a perfectly-reconstructing zero-phase one-third-octave filter and $F_{\text{av}}\{\cdot\}$ denotes a time averaging over 10 ms.

### 5.3 Merged interpolated peaks and residual ARIRs

The final interpolated ARIR $d(t)$ at the listener position is obtained by summing up the interpolated peaks $d_p(t)$ and the interpolated residual ARIRs $d_r(t)$ (cf. Fig. 10). An additional rotation to the listener's head orientation $\psi_d$ enables the convolution-based spatial auralization of any single-channel signal $s_{\text{in}}(t)$ at a desired listener perspective in the captured acoustic environment,

$$s_d(t) = (s_{\text{in}} \times d_\psi)(t), \quad (42)$$

$$\text{with} \quad d_\psi(t) = R_{\text{sp}}^{(N)}(\psi_d)[d_p(t) + d_r(t)], \quad (43)$$

where $s_d(t)$ is the Ambisonic signal auralized at the variable listener perspective and $R_{\text{sp}}^{(N)}(\psi_d)$ is an $N$th-order spherical harmonics rotation matrix as described in Section 2.2 (6) that rotates the azimuth and zenith of the interpolated ARIR according to the listener's head orientation.

### 5.4 Real-time interpolation from intermediate fine-meshed ARIRs

Due to non-negligible computational costs, a real-time capable algorithm for all the steps in Figure 8 would be quite demanding, even if all position-independent operations are carried out in advance. This section therefore presents a real-time capable system based on an offline-rendered, fine-meshed grid of interpolated ARIRs and a simplified interpolation thereof at the expense of increased memory requirements.

Initially, the coarse ARIR grid is interpolated to a sufficiently fine ARIR grid using the proposed algorithm. We propose a grid spacing of e.g. $r = 0.25$ m, where perceptible but moderate directional and timbral differences appear between neighboring interpolated ARIRs. The choice of the grid spacing is a trade-off between a preferably fine ARIR resolution and limited memory requirements. The real-time 6DoF ARIR rendering can then be achieved by a low-cost interpolation of the pre-computed ARIR grid. We propose to interpolate the three closest pre-computed ARIRs $d_i(t)$, where early ARIR parts are linearly interpolated and the late ARIR part, e.g. for $t > 100$ ms, is solely taken from the nearest neighbor.

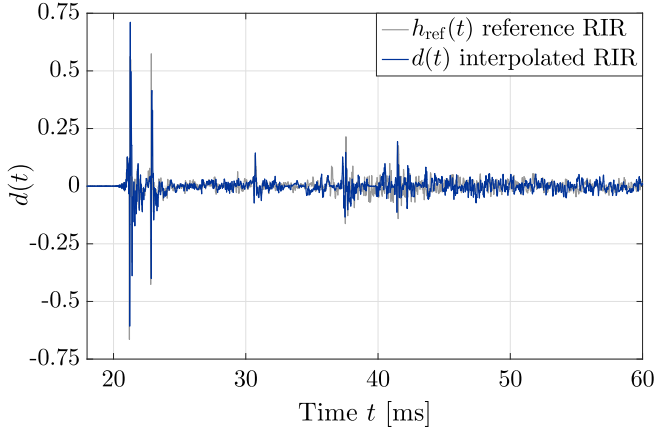

**Figure 10.** Interpolated ARIR $d(t)$ in comparison to a reference RIR $h_{\text{ref}}(t)$ measured at the target listener perspective.

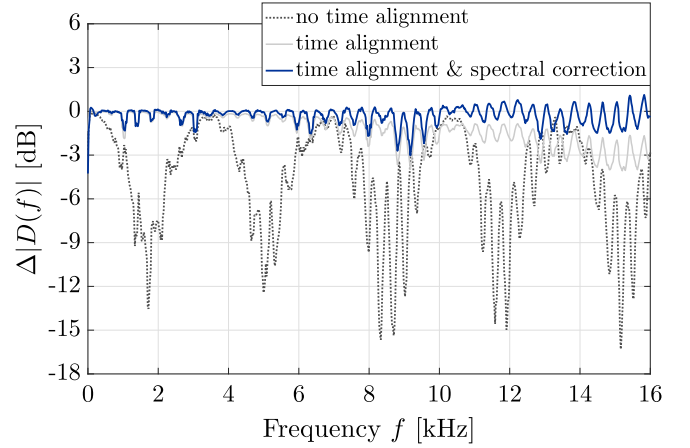

**Figure 11.** Spectral differences $\Delta|D(f)|$ between the interpolated ARIR and the reference ARIR spectrum for different real-time interpolation methods.

The neighboring ARIRs of the re-fined grid predominantly contain TOA shifts and nearly no directional variation. And yet, pure linear interpolation would not work. For instance, the maximum possible TDOA between neighboring grid points with $r = 0.25$ m is $\Delta T_{\text{DS,max}} \approx 0.73$ ms and linear ARIR interpolation could yield comb filtering with a clearly noticeable first notch at a frequency as low as $f \approx 700$ Hz. To reduce the most dominant comb filter that is usually induced by the temporal deviation of the direct-sound peaks, we propose a time alignment of the pre-computed ARIRs $d_i(t)$ compensating for the direct-sound peak TOA differences. To this end, the early ARIRs are shifted to a common direct-sound peak TOA,

$$T_{\text{DS}} = \sum_{i=1}^{3} g_i(\boldsymbol{x}_{\text{d}}) \, T_{\text{DS},i}, \quad (44)$$

before linear interpolation,

$$\boldsymbol{d}(t) = \sum_{i=1}^{3} g_i(\boldsymbol{x}_{\text{d}}) \, \boldsymbol{d}_i(t - (T_{\text{DS}} - T_{\text{DS},i})), \quad (45)$$

where $T_{\text{DS},i}$ are the direct-sound TOAs and $g_i(\boldsymbol{x}_{\text{d}})$ are the distance-dependent amplitude weights (1) of the pre-computed ARIR triplet $\boldsymbol{d}_i(t)$. Figure 11 exemplarily shows the influence of the time alignment on the spectral difference $\Delta|D(f)|$ between the magnitude spectrum of the interpolated ARIR $\boldsymbol{d}(t)$ and the averaged magnitude spectrum of each $\boldsymbol{d}_i(t)$, $i \in \{1, 2, 3\}$.

As linear interpolation of time-aligned ARIRs may still cause small undesirable artifacts such as spectral ripple or roll-off (cf. Fig. 11 for $f > 10$ kHz), an additional spectral correction of the temporal envelope in one-third-octave bands as described in Section 5.2 (39) is reasonable and also displayed in Figure 11.

When highly directional sources are involved, early reflection peaks could exceed the direct sound level and thus direct-sound time alignment only would not prevent distinct comb filtering. While unproblematic in this paper, an alternative approach could consider adapting real-time dynamic time warping as in the BRIR interpolation approach of [26]. It improves interpolation of closely spaced ARIRs by avoiding small differences in timing. Another feasible alternative could separate real-time rendering of fine-mesh-interpolated ARIR residuals from rendering of the $M$ most distinct matching-peak sound events of the current, coarse ARIR-triplet. These $M$ few, short sound-event segments can still be efficiently convolved, extrapolated (time, level, rotation), and linearly interpolated while rendering in real time.

### 5.5 Time-variant overlap-add convolution

We implemented the real-time 6DoF auralization using a STFT-based convolution in frames of $T_{\text{s}} = 1024$ samples with an ARIR update after each frame. To this end, we split the current interpolated ARIR $\boldsymbol{d}(t)$ and the recent equal-length section of the single-channel input signal $s_{\text{in}}(t)$ into $N_{\text{s}}$ non-overlapping segments of length $T_{\text{s}}$ and compute a $2T_s$-point DFT $\mathcal{F}\{\cdot\}$ of each segment,

$$\begin{aligned}\hat{\boldsymbol{d}}^{(k)}(f) &= \mathcal{F}\{\boldsymbol{d}(\tau)\}, \quad kT_{\text{s}} \leq \tau < (k+1)T_{\text{s}},\\ \hat{s}_{\text{in}}^{(k)}(f) &= \mathcal{F}\{s_{\text{in}}(t-\tau)\}, \quad k \in \{0, 1, \ldots, N_{\text{s}}-1\}.\end{aligned} \quad (46)$$

The Ambisonic output signal $\boldsymbol{s}_{\text{d}}(t)$ is then updated after each $T_{\text{s}}$ period by overlap-add of the convolved Ambisonic output frame $\boldsymbol{s}_{\text{out}}(t)$,

$$\boldsymbol{s}_{\text{d}}(t+\tau) \leftarrow \boldsymbol{s}_{\text{d}}(t+\tau) + \boldsymbol{s}_{\text{out}}(t+\tau), \quad 0 \leq \tau < 2T_{\text{s}}$$

with $\quad \boldsymbol{s}_{\text{out}}(t+\tau) = \mathcal{F}^{-1}\left\{\sum_k \hat{s}_{\text{in}}^{(k)}(f) \cdot \hat{\boldsymbol{d}}^{(k)}(f)\right\}, \quad (47)$

where $\mathcal{F}^{-1}\{\cdot\}$ denotes a $2T_{\text{s}}$-point inverse DFT.

**Table 1.** Configurations of presented stimuli.

| Label | ARIR type | $M$: Number of matched peaks | Extrapolation of residual ARIRs |
|---|---|---|---|
| HR | Measured | – | – |
| S | Simulated | – | – |
| A | Interpolated | 0 | ✗ |
| B | Interpolated | 1 (DS) | ✗ |
| C | Interpolated | 1 (DS) | ✓ |
| D | Interpolated | 11 (DS + 10 ERs) | ✗ |
| E | Interpolated | 11 (DS + 10 ERs) | ✓ |

# 6 Listening experiment evaluation

To finalize our investigations, we carried out a listening experiment to evaluate the effectiveness of the various system components on perceptual features, such as localization and sound coloration. Specifically, the influence of separated versus jointly localized early sound events as well as the extrapolation of residual ARIRs prior to their interpolation was of particular interest. We used the most basic interpolation method $A$, the simple distance-weighted linear combination of the closest ARIR triplet (2), as comparison to more complex configurations $B\ldots E$. As listed in Table 1, configurations $B, C$ are characterized by a separate interpolation of the direct-sound peak (DS) only and $D, E$ by interpolation of eleven matched peaks (DS + 10 early reflections, ERs) (cf. Sect. 5.1). Furthermore, $B, D$ apply no extrapolation of the residual ARIRs, i.e. the unmodified residual ARIRs are linearly interpolated by distance-weighted superposition as done in $A$, whereas $C, E$ additionally extrapolate the residual ARIRs before interpolation (cf. Sect. 3.2).

## 6.1 Experiment setup

The variable-perspective ARIRs for convolution-based auralization were interpolated from a horizontal first-order ARIR grid with equidistant $r = 2$ m spacing. Besides a measured dataset[2] with 30 first-order ARIR positions recorded in the IEM CUBE ($T_{60} \approx 0.65$ s), we simulated a first-order ARIR grid using the default settings of the "medium room" of the Matlab-Toolbox *MCRoomSim*[3] [47], however with reduced room dimensions (14 × 10 × 4.1 m) and room absorption (80% of the default value). Both the measured and simulated ARIR grids were directionally enhanced to 5th-order by ASDM. For the rendering of stimuli[4], we used the real-time capable method described in Section 5.4. To this end, offline interpolation was done to a fine-meshed 5th-order ARIR grid with $r = 0.2$ m spacing with each configuration $A\ldots E$. On this re-fined grids, real-time capable interpolation was applied including the time alignment and spectral correction described above.

The stimuli were presented to the listeners with head-tracked Beyerdynamic DT-770 Pro headphones in an anechoic room. For the interactive head rotation and binauralization of the 5th-order Ambisonic stimuli, we used the *SceneRotator* and *BinauralDecoder* [48, 49] ([38], Ch. 4) of the IEM plugin suite[5]. A graphical user interface for multi-stimulus trials, similar to MUSHRA testing [50], was used for evaluation, which enabled arbitrary repetition and pair comparisons of the presented stimuli. All stimuli were presented in randomized order.

The listening experiment consisted of three parts. In the first part, a measured ARIR triplet was used to interpolate an ARIR with each configuration $A\ldots E$ at a static listener position inside the ARIR perspectives. The listeners were asked to rate the,

1. perceived similarity of localization, in particular the direction and distance impression of the signal (stimulus: *speech*),
2. perceived similarity of sound coloration (*pink noise*),

with regard to a given reference stimulus. The reference was rendered by convolution with an ARIR measured at the same listener perspective. To evaluate the reliability of individual ratings, the reference was also added as hidden reference (HR).

In the second and third part, dynamic listener perspectives were evaluated (cf. Fig. 12). To this end, we rendered variable ARIRs at a moving listener perspective offline with an update every 1024 samples and auralized the output signal using the real-time capable convolution described in Section 5.5. The second part auralized a virtual walk on a straight line through the measured room (IEM CUBE) based on the recorded ARIR grid. No reference signal was available in this part.

In the third part, we used the simulated ARIR grid to auralize the stimuli $A\ldots E$ of a virtual walk on a circular path around the source position. Additionally, we utilized the room simulation to simulate ARIRs in steps of 2 cm on the circular path. This enabled to render a simulated reference signal $S$ for comparison. The listeners were asked to rate the,

1. perceived smoothness of localization, in particular of the direction and distance impression (stimulus: *speech*),
2. perceived smoothness of sound coloration (*pink noise*).

## 6.2 Listening experiment results

19 experienced listeners with an average age of 27 took part at the listening experiment. They needed 23 min on average to evaluate the presented stimuli. Responses of one listener were unreliable and excluded from the evaluation as they contained a standard deviation that was 2.5 times larger than the standard deviation of other listeners' responses for the hidden reference. The similarity of the results of Part 2 and 3 allowed to pool and commonly evaluate the ratings of dynamic listener perspectives (note that the sample size of $S$ is half the sample size of $A\ldots E$, since no reference was available in Part 2). Figures 13 and 14 show the median and 95% confidence intervals of the ratings.

---

[2] https://phaidra.kug.ac.at/o:104435
[3] https://github.com/Andronicus1000/MCRoomSim
[4] All presented stimuli are available (binaural and in 5th-order Ambisonics) on https://phaidra.kug.ac.at/o:104443
[5] https://plugins.iem.at/

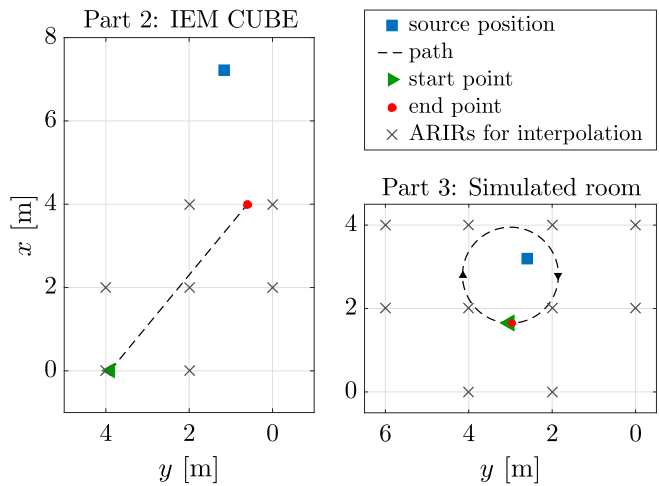

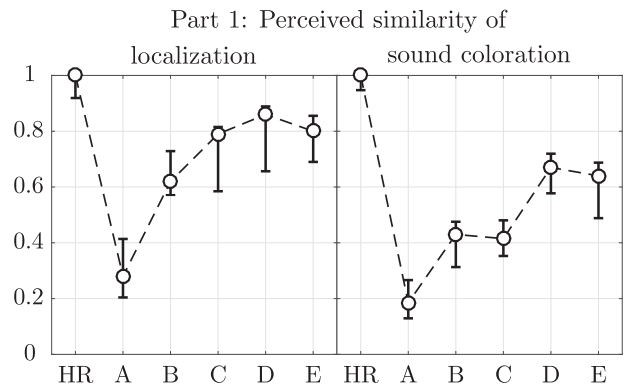

Figure 13. Median and 95% confidence intervals of the ratings for a static listener perspective with regard to a measured reference.

Figure 12. Measured (Part 2) and simulated (Part 3) 5th-order ARIR grids used for the variable-perspective ARIR interpolation at a dynamic listener perspective.

For evaluation, we computed univariate ANOVAs with repeated measures for more than two paired samples. The significance between presented stimuli was evaluated via post hoc Bonferroni tests.

### 6.2.1 Static listener perspective

*Perceived similarity of localization with regard to a given reference*: Most listeners recognized the hidden reference (HR) to be identical with the reference stimulus (median = 1). The lowest-rated condition $A$, the linearly interpolated ARIR triplet, significantly differs from all other stimuli ($p < 0.001$), whereas no significant differences are found between $B$, $C$, $D$, $E$ ($p = 0.25$).

*Perceived similarity of coloration with regard to a given reference:* Most listeners recognized the hidden reference (HR) to be identical with the reference stimulus (median = 1). The ratings of $A$ are significantly lower than all other stimuli ($p < 0.001$). Furthermore, stimuli $D$, $E$, characterized by a higher number of interpolated matched peaks, were rated significantly better than $B$, $C$ ($p < 0.001$), whereas there is no significant difference between $B$ and $C$ or between $D$ and $E$ ($p = 1$).

### 6.2.2 Variable listener perspective

*Perceived smoothness of localization:* The simulated signal $S$ is rated highest (median = 0.83, mean = 0.77), closely followed by the full interpolation system $E$ (median = 0.79, mean = 0.76). The lowest-rated condition $A$ significantly differs from all other stimuli ($p < 0.001$). However, no significant difference is observed to distinguish between the simulated reference $S$ and the configurations $B$, $C$, $D$, $E$ ($p = 0.36$).

*Perceived smoothness of sound coloration:* The simulated signal $S$ is rated highest (median = 0.80, mean = 0.75), closely followed by the full interpolation system $E$ (median = 0.75, mean = 0.75). Significant differences are determined between $S$, $E$ and $A$, $B$, $D$ ($p < 0.001$), whereas no significant differences are observed within $S$, $C$, $E$ ($p = 0.24$) and within $A$, $B$ ($p = 1$).

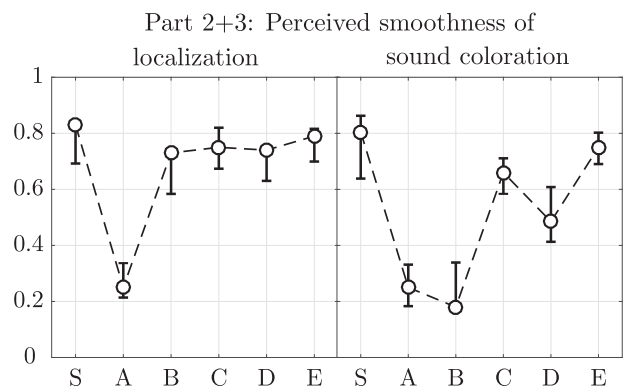

Figure 14. Median and 95% confidence intervals of the ratings for a variable listener perspective.

### 6.3 Discussion

The listening experiment proved that the proposed ARIR interpolation system distinctly outperforms the most basic, linear interpolation of the closest ARIR triplet for both static and dynamic listener perspectives.

Compared to the separate interpolation of the matched direct-sound peak only ($B$, $C$), an additional effort to jointly localize and interpolate the ten most prominent early reflections ($D$, $E$) is clearly superior in terms of the perceived sound coloration for static perspectives (cf. Fig. 13). However, an additional extrapolation of the residual ARIRs yields no further improvement when interpolating sufficiently many matched ARIR peaks ($E$) and thus can be omitted for static perspectives.

The evaluation of a time-variant, moving perspective shows no notable differences of the perceived smoothness of localization concerning any of the proposed configurations, except for the poorly-rated, plain linear interpolation ($A$), which was perceived to be largely static, though interrupted by abrupt changes in location and coloration. However, the joint localization and interpolation of the ten most prominent matched early reflections ($D$, $E$) was clearly superior concerning the perceived smoothness of sound coloration for variable perspectives. This matches with the observation that disturbing comb filter fluctuations can be distinctly reduced due to the position-dependent rendering of jointly localized prominent ARIR peaks. Other than for static listener perspectives, extrapolating the residual ARIRs before interpolation ($C$, $E$) clearly improves the perceived smoothness of sound coloration for variable perspectives, where the full system for interpolation ($E$) was rated similarly high as the simulated reference stimulus ($S$). This justifies applicability of the residual extrapolation based on the simplistic single-ARIR localization of instantaneous sound events that could not be jointly localized. Even though this localization technique is less precise, it still grants a clear improvement in terms of a smooth transition of variable, position-dependent ARIR comb filters.

The ratings we obtained on localization do not appear quite as selective as those obtained for sound coloration in both of the scenarios evaluated. For little coloration in the static scene, many geometrically precisely rendered sound events were important ($D$, $E$), and for little coloration in the dynamic scene, the extrapolation of the residuals ($C$, $E$) was superior to plain linear interpolation. A DTW-based approach [24, 25, 36] extended to higher-order Ambisonics might appear worth examining as an alternative. While its localization might not be treated as accurately, its time-aligning warping targets an interpolation avoiding coloration in low-order channels.

As often, the perception of artifacts and sound coloration in the investigated application appears to be less critical for speech signals compared to music or noise signals. Moreover, differences between the algorithmic configurations vanish in strongly reverberant environments with long and highly diffuse impulse response tails, which appear to be uncritical. Therefore, the proposed system proves to be relevant especially in rather weakly or medium reverberant environments and in applications with music.

## 7 Conclusion

In this paper, we presented a parametric Ambisonic room impulse response (ARIR) interpolation system that achieves convolution-based auralization of a variable listener perspective when a grid of measured or simulated, spatially distributed ARIRs is available to describe the acoustic environment of interest. To improve the directivity and naturalness of spatial reproduction, it moreover considers an optional enhancement of the directional resolution when using first-order ARIR grids. The fundamental auralization principle is to parametrically extrapolate the perspectives of a local ARIR triplet to the desired listener perspective preceding the linear interpolation of the three ARIRs. In particular, we detect distinct early ARIR peaks that can be assigned to common sound events, such as direct-sound source and image sources, and we jointly localize corresponding sound-event positions in the ARIR triplet around the variable listener position. What cannot be resolved in this scheme remains in the residual ARIRs, for which a simplistic instantaneous sound-event localization is applied based on single-perspective information. With such estimated sound-event positions, the three ARIRs can be extrapolated in time segments that preserve the temporal and directional context as much as possible when extrapolated by re-assigning direction, time, and level of the segments with regard to what would be observed at the desired listener perspective. Finally, a linear, distance-weighted interpolation is employed that restores the spectral and temporal RMS level of the resulting ARIR to compensate for varying interference. We furthermore introduced a real-time capable system based on a simplified interpolation from an offline-interpolated, fine-meshed ARIR grid.

Conclusively, the proposed ARIR interpolation system was evaluated at different algorithmic settings in a listening experiment, which proved a distinctly better quality in terms of localization and sound coloration compared to a basic, distance-weighted linear interpolation. But also the proposed steps that appear to require extra efforts could be supported and justified by the results of the experiment. Not as much in terms of localization but more in terms of a minimal sound coloration, the extrapolation of the residual early sound events was rated superior when the listener perspective is time-variant. For a static listener, raising the number of jointly localized early sound events from one to eleven was rated superior.

As the techniques discussed in our evaluation are not the only ways of obtaining interpolated ARIRs, the underlying ARIR measurements[2] and the resulting processed binaural and Ambisonic experimental stimuli[4] are made available online for reference and to encourage comparative studies.

The authors filed priority rights for the method described herein [51].

## Conflict of interest

Author declared no conflict of interests.